\documentstyle[twocolumn,aps,epsf]{revtex}
\def\lamchi{{\Lambda_\chi}}

\def\bra#1{{\langle#1\vert}}
\def\ket#1{{\vert#1\rangle}}
\def\sst#1{{\scriptscriptstyle #1}}
\def\alrd{{A_\sst{\gamma}}}
\def\alre{{A_e}}

\begin{document}

\twocolumn[\hsize\textwidth\columnwidth\hsize\csname
@twocolumnfalse\endcsname

\title{ Parity Violating Photoproduction of $\pi^\pm$ on the $\Delta$
Resonance}
\author{Shi-Lin Zhu$^{a,b}$, C. M. Maekawa$^{a}$,   B.R. Holstein$^{c}$,
and M. J. Ramsey-Musolf$^{a,b,d}$\\
$^a$ Kellogg Radiation Laboratory 106-38, California Institute   of
Technology, Pasadena, CA 91125\\
$^b$ Department of Physics, University of Connecticut,   Storrs, CT 06269 \\
$^c$ Department of Physics, University of Massachusetts,   Amherst, MA
01003 \\
$^d$ Theory Group, Thomas Jefferson National Accelerator Facility, Newport
News, VA 23606 }
\maketitle

\begin{abstract}
We analyze the real photon asymmetry $A^\pm_\gamma$ for the parity
violating (PV)
$\pi^\pm$ production on the $\Delta$ resonance via the   reactions
${\vec \gamma} + p \to \Delta^+ \to \pi^+ + n$ and
${\vec \gamma} + d \to \Delta^0 + p \to \pi^-  +  p + p$.
This   asymmetry is nonvanishing due to a PV $\gamma N \Delta$ coupling
constant, $d^\pm_\Delta$.
We argue that an experimental determination of   this coupling
would be of interest for hadron dynamics,
possibly shedding   light on   the S-wave/P-wave puzzle in the hyperon
nonleptonic decays
and   the violation of Hara's theorem in weak radiative hyperon decays.


\medskip
PACS Indices: 11.30.Rd, 13.88.+e, 13.60.Le
\end{abstract}
\vspace{0.3cm}
]

\pagenumbering{arabic}

Despite years of scrutiny, the strangeness-changing ($\Delta S=1$) hadronic
weak interaction
(HWI) continues to  confront hadron structure physicists with a number of
largely
unsolved   puzzles: {\em e.g.}, the origin   of the $\Delta I=1/2$ rule,
the S-wave/P-wave problem in hyperon nonleptonic   weak decays, and the
surprisingly large violation of SU(3) symmetry in hyperon
weak radiative decays. These puzzles involve the breakdown of symmetries
associated with
light quark and gluon strong interactions when applied to hadronic weak
observables.
Indeed, hadronic weak matrix elements involve a complex interplay of strong
and weak interactions,
and the degree to which QCD-based symmetries are relevant in this context
remains an open question
for the field.

Our understanding of the strangeness-conserving ($\Delta S=0$) HWI
remains no less opaque, and similar questions about the applicability of
strong interaction symmetries arise in this case.  Recently, there has been
resurgence of
interest in studying the $\Delta S=0$ HWI via measurements of
parity-violating (PV),
strangeness-conserivng observables \cite{parity}. The use of parity
violation allows one to filter
out the $\Delta S=0$ HWI from  much larger, strangeness-conserving strong and
electromagnetic effects. While these PV experiments focus on nucleonic
systems, there have yet to be
any studies of  strangeness-conserving weak {\em  transitions}  between the
nucleon ($N$) and other
$qqq$ states.  In this note, we propose such a probe of the $\Delta S=0$ HWI
involving the nucleon and its lightest spin-$\frac{3}{2}$   partner,
the   $\Delta(1230)$ resonance:   $\alrd^\pm$, the PV $\pi^\pm$
photoproduction
asymmetry at the $\Delta$   resonance using polarized photons.

In the limit where the number of quark and gluon colors ($N_c$) becomes large,
the $N$ and $\Delta$ form a degenerate multiplet under
spin-flavor SU(4) symmetry, making this system an interesting
window on   the hadronic dynamics of light $qqq$ hadrons.
Measurements of the   parity-conserving   (PC) electromagnetic (EM)
$N\to\Delta$
transition amplitudes (Fig. 1a) have   challenged hadron structure theorists,
as the experimental results differ   substantially from both
lattice QCD and QCD-inspired model predictions \cite{hhk}.
A determination of the axial vector neutral current   $N\to\Delta$
amplitude via PV pion electroproduction planned at Jefferson
Laboratory   \cite{proposal}  could, in principle, shed additional
light on the   present shortcomings   of hadron structure theory in
this context.  In a separate communication, we show that
a parallel measurement   of
$\alrd^\pm$ could sharpen the theoretical interpretation of the PV
electroproduction  asymmetry, $A_e$ \cite{long}. Here, we concentrate on
the hadron structure implications of $\alrd^\pm$ -- whose measurement
also appears to   be feasible at  Jefferson Lab\cite{krl} -- and show
how it could provide new insight  into the physics of the poorly understood
HWI
in both the $\Delta S=0$ and $\Delta S=1$ sectors.

The relationship between $\alrd$ and its $\Delta S=1$ partners
follows from the QCD-based symmetry properties -- chiral, flavor SU(N),
large $N_c$ -- of the
relevant transition matrix  elements. Chiral perturbation theory ($\chi$PT)
provides a natural
and systematic framework for applying these symmetries to hadronic observables.
At leading order in the chiral expansion,
$\alrd^\pm$ is dominated by the
PV $\Delta\to N\gamma$ electric dipole (E1) amplitude, which is the $\Delta
S=0$, SU(4)
analog of the   SU(3)-forbidden E1 amplitude responsible
for PV $\Delta S=1$ hyperon weak radiative decays. In the limit of exact
SU(3) symmetry, the
asymmetry paraemters $\alpha^{BB'}$ associated with the latter  must
vanish, a result known as Hara's
theorem\cite{hara}. The non-zero splitting between the strange and light
quark masses breaks the
SU(3) symmetry, leading one to expect $\alpha^{BB'}\sim m_s/m_N\sim 0.15$,
where $m_s$ is
the strange quark mass. Experimentally, however,
one finds \cite{pdg,prl}
$\alpha^{\Sigma^+ p}  =  -0.76\pm 0.08, \qquad
\alpha^{\Xi^0\Sigma^0}  =  -0.63\pm0.09$,
roughly four to five times the naive expectation.

In an analogous manner, the PV $\Delta\to N\gamma$ E1
transition   -- and thus the resonant  asymmetry $\alrd^\pm$   -- must vanish
in the limit of exact SU(4) symmetry. It is   natural to ask,   then,
whether the dynamics underlying the   surprisingly large
SU(3)-violation in the $\Delta S=1$ sector also produce
enhanced SU(4)-breaking in the $\Delta S=0$ weak radiative
transition. Using a   specific mechanism for both
the $\Delta S=1$ and $\Delta S=0$ transitions,   we show
that   one might reasonably expect enhancements to be
generated in both sectors by   similar   dynamics. More
generally, a measurement of $\alrd^\pm$ could help
determine   whether   anomalously large symmetry breaking
is a general feature of low-energy   hadronic weak   interactions
or simply a peculiarity of $\Delta S=1$ hyperon decays.


Our result for the $\alrd^\pm$ leading to these conclusions
has the simple form:
\begin{equation}
\label{eq:photon1}
\alrd^\pm \approx -{2d^\pm_\Delta\over C_3^V}{M_N\over\lamchi}+\cdots
\end{equation}
where $\alrd^\pm$ is the PV  asymmetry on the $\Delta$   resonance,
$\lamchi=4\pi F_\pi\sim 1$ GeV is the scale of chiral symmetry   breaking,
$C_3^V\sim 2$ is the dominant $N\to \Delta$ vector transition
form factor (Fig. 1a) as defined in Ref. \cite{adl},
$d^\pm_\Delta$ is a low-energy constant (LEC) characterizing
the PV   $\gamma N\Delta$ coupling in Fig. 1b, and the
ellipses denote non-resonant,   higher   order chiral,
and $1/M_N$ corrections.  Several features of $\alrd^\pm$   merit comment:

\medskip   \noindent (i) The photoproduction asymmetry coincides
with the $q^2\to 0$ limit   of $\alre$. The nonvanishing   $\alre(q^2=0)$
results from $\gamma$-exchange between the electron and   hadronic target,
with the $\gamma$-hadron coupling  given in Fig. (1b)    corresponding to
matrix elements of the transverse electric multipole
operator   ${\hat T}^\sst{E}_{J=1\lambda}$\cite{henley}. The constraints
of current conservation, as   expressed in Siegert's theorem
\cite{Siegert,friar},
allow one to rewrite   matrix elements of this operator in terms of the
corresponding charge density   matrix elements:
\begin{eqnarray}\nonumber   
\bra{f}{\hat T}^\sst{E}_{J=1\lambda}\ket{i} =
-{\sqrt{2}\over 3}\omega\bra{f}
\int d^3x\ xY_{1\lambda}(\Omega){\hat\rho}(x)\ket{i} +{\cal O}(q^2) \; ,
\end{eqnarray}
where the $\omega$ is the energy difference between the
initial and final   states and vanishes in the $N_c\to\infty$
limit wherein the $N$ and $\Delta$   become degenerate\cite{dashen}.
The leading term is $q^2$-independent and proportional to $\omega$
times the   electric dipole matrix element. Up to numerical factors,
this matrix element   is simply $d_\Delta^\pm/\Lambda_\chi$. The
remaining   terms of ${\cal O}(q^2)$ and higher contain matrix
elements of the anapole   operator \cite{hax89,mh}, which do not
contribute to $\alrd^\pm$   but  generate contributions to $\alre$
that vanish at the photon point.

\medskip   \noindent (ii) In the context of
$\chi$PT, one expects   the \lq\lq natural" scale for SU(4)
breaking effects associated with   $d^\pm_\Delta$ to
be $\sim\ {\mbox{few}}\times   10^{-8}$ (see below).
However, the magnitude of observed $\Delta S=1$ PV
asymmetries   suggest that   $d_\Delta^\pm$ could be
significantly enhanced over this scale,   yielding a
potentially relatively large   real photon asymmetry
$\alrd^\pm\sim $   ${\mbox{few}}\times 10^{-6}$, which
would be accessible   using polarized photons at Jefferson Lab.

In performing a consistent derivation of the photoproduction
asymmetry, one   must   consider  all contributions to the PV
amplitudes through a given chiral order.   However, while one
may readily identify the formal   chiral order of various
contributions to $A^\pm_\gamma$,   the physical significance
of chiral counting is complicated   by the dominance of
the   $\Delta$ intermediate state at resonant kinematics.
In particular, we do not   include various   non-resonant
background contributions, even though some may be formally
of   lower chiral order than those involving the $\Delta$
intermediate state (see,   {\em e.g.} the studies of PV
threshold $\pi$ production in Refs.   \cite{CJ01,zhu2,CJ02}).
The reason is that for resonant kinematics, the
contribution of the   $\Delta$ is enhanced relative to the
non-resonant (NR) background   contributions by
$   \sigma^\Delta/\sigma^{NR}
\sim \left(2M_\Delta/\Gamma_\Delta\right)^4\sim   2\times 10^4$.
This enhancement factor more than compensates for   the relative
chiral orders of the $\Delta$ and NR   contributions. Indeed,
from a blind application of chiral power   counting to $A^\pm_\gamma$,
one might erroneously   truncate the chiral expansion at ${\cal O}(p)$,
retaining only the NR   background contributions to the
resonant asymmetry. Hence,  we use   chiral power counting
as a means of organizing various resonant   contributions
instead of using it to delineate the relative importance
of resonant and NR amplitudes.

To that end, we employ  heavy baryon chiral
perturbation theory (HB$\chi$PT)   \cite{j1,ijmpe}   and
adopt the $p$-counting scheme, where $p$ denotes  a
small external   momentum or mass or the photon field. The
leading PV $\Delta\to N\gamma$   transition operator is
then ${\cal O}(p^2)$ \cite{zhu,CJ01}:
\begin{equation}   \label{eq:Siegert2}
{\cal L}^{\Delta N\gamma}_\sst{PV} = i{e \over\lamchi}
[ d^+_\Delta {\bar   \Delta^+}_\mu\gamma_\lambda p   +
 d^-_\Delta {\bar \Delta^0}_\mu\gamma_\lambda n ]
F^{\mu\lambda}+{\mbox{h.c.}}
\end{equation}
and we truncate our expansions of $d_\Delta^\pm$ at this order.
The ${\cal O}(p^3)$ corrections -- including loop effects -- are generally
small and can   be found in \cite{long}.
\footnote{Another possible
 resonant subleading   correction comes from PV $\pi N \Delta$ D-Wave
 interaction.   However, a careful analysis shows it does not
 contribute to   the total real photon asymmetry \cite{long}.}

Separate determinations of $d_\Delta^+$ and $d_\Delta^-$ could
be achieved   using proton and deuterium targets, respectively.
In the latter case,   the resonant $\pi^-$ production process
${\vec \gamma} + d \to \pi^-  + p + p$   is dominated by the
subprocess   ${\vec \gamma} + n \to \Delta^0  \to \pi^-  +  p$
since two body   meson exchange currents are always higher order
due to   the presence of an additional loop \cite{review}.   The
asymmetry derivation is   the same for $\pi^\pm$ so we take $\pi^+$
resonant production as a specific example.   Defining the kinematic
variables as
\begin{equation}   \label{eq16}   {\vec \gamma} \left( q\right)
+p\left( p\right) \rightarrow   \Delta^+ \left( p_\Delta \right)
\rightarrow   n \left( p^{\prime }\right) +\pi^+ \left( k \right) ,
\end{equation}
we have in the laboratory frame
$s=\left( k+p\right) ^2, q=p_\Delta -p,   p_\Delta =p^{\prime }
+p_\pi $   where ${\bf p}=0$ and $p_\Delta^2= m_\Delta^2$.
The PV asymmetry is defined as
$ A_{\gamma}=(N_+ - N_-)/(N_++N_-)$,   where $N_{+}$ ($N_{-}$) is
the number of detected  $\pi^+$ produced via   the
reaction (\ref{eq16})   for a beam of left (right) handed circularly
polarized photons.

To compute $\alrd^+$, we require the PC and PV response functions,
generated   by photon helicity ($h$) amplitudes $M_{PC}^h$ (Fig. 1a)
and $M_{PV}^h$   (Fig. 1b),   respectively:
$W_{PC}=|M^+_{PC}|^2 +|M^-_{PC}|^2$
and   $W_{PV}=2 \mbox{Re}
\left( M^{+\ast}_{PC} M^+_{PV} -M^{-\ast}_{PC} M^-_{PV}   \right)$.
A straightforward calculation leads to
\begin{eqnarray}\nonumber
 W_{PC}&=&{32\pi \alpha\over 9}  S_\Delta^2
\left({g_{\pi N \Delta} C_3^V \over F_\pi }\right)^2
|{\bf q}|^2 |{\bf k}|^2 ({5\over 3} -\cos ^2 \theta )\\  \nonumber
 W_{PV}&=&-{64\pi \alpha\over 9} m_N S_\Delta^2
\left({g_{\pi N \Delta}  \over F_\pi }\right)^2
{d^+_\Delta\over \Lambda_\chi} C_3^V \\
&&\times |{\bf q}|^2 |{\bf k}|^2 ({5\over 3} -\cos ^2 \theta )
\end{eqnarray}
where $S_\Delta =\left( s-m_\Delta^2 +\Gamma^2_\Delta/4\right)^{-1}$
and $g_{\pi N \Delta}$ is the $\pi N \Delta$ strong coupling constant.
The asymmetry is given by the ratio $W_{PV}/W_{PC}$, yielding the
result   in Eq. (\ref{eq:photon1}). It is interesting   to note that
the leading asymmetry in HB$\chi$PT is isotropic.

In order to assess the potential size of this contribution
to the   asymmetry, we need to estimate the size of
$d_\Delta^\pm.$ Because  LEC's   such as $d_\Delta^\pm$ are
governed in part by short-distance   ($r>1/\Lambda_\chi$)
strong interactions, they are difficult to compute from
first principles in   QCD.   A standard alternative is to
employ   \lq\lq naive dimensional analysis" (NDA) \cite{georgi},
according to which   effective   weak interaction operators should
scale as \cite{zhu2}
\begin{equation}   \left({{\bar\psi}\psi\over \lamchi F_\pi^2}\right)^k
\left({\pi\over   F_\pi}\right)^\ell   \left({D_\mu\over\lamchi}\right)^m
\times (\lamchi F_\pi)^2 \times g_\pi\ \ \ ,
\end{equation}
where $   g_\pi\sim {G_F F_\pi^2 / 2\sqrt{2}}\sim 5\times 10^{-8} $
is the scale of weak charged current hadronic processes.
The SU(4)-violating E1   operator of Eq. (\ref{eq:Siegert2})
corresponds to $k=1$, $\ell=0$ and   $m=2$ and should scale
as $g_\pi/\lamchi$, so that $d_\Delta^\pm\sim g_\pi$.
A more detailed consideration of hadron dynamics, however,
suggests   that $|d_\Delta^\pm|$ may be considerably larger.

In the purely mesonic sector of $\chi$PT, one knows that low energy
constants   are well reproduced by assuming resonance saturation for
the short distance   physics. In the baryon sector, a particularly
intriguing application of   resonance saturation involves the electric
dipole transitions responsible for the anomalously large asymmetries
$\alpha^{BB'}$ discussed above. The theoretical challenge has been to
account for these
enhanced values of   $\alpha^{BB'}$ in a manner consistent with
the corresponding nonleptonic   pion decay rates.

While a number of approaches have been attempted, the
inclusion of   $\frac{1}{2}^{-}$   resonances as in Fig. 2a
appears to go furtherst in enhancing the   theoretical
predictions for the asymmetries while simultaneously
resolving the S-wave/P-wave problem in the pion decay
channel \cite{resonance,resonance1}.   Here, the pseudoscalar,
nonleptonic weak interaction ${\cal H}_W^{PV}$   mixes states of
the same spin and opposite parity into the initial and
final baryon states, and if $\frac{1}{2}^{-}$ resonance
saturation is indeed   the correct   explanation for the
enhanced $\Delta S=1$ PV radiative asymmetries,   then
one might naturally expect a similar mechanism to play
an   important role in the $\Delta S=0$ PV electric dipole
transition   examined in this paper. The relevant diagrams
appear in Figs. 2b and   2c, where the intermediate states
have $J^\pi=\frac{1}{2}^-$ and   $\frac{3}{2}^-$,
respectively, and where the $\gamma$ vertex brings
about the change in spin.   In using this picture, we note that:

\medskip   \noindent (i) At present, one has detailed
information on the   $\frac{1}{2}^{-}\leftrightarrow\frac{1}{2}^{+}$
$\Delta S=1$ amplitudes
from   fits to the S-wave $B\pi$ decay channel. In contrast, {\it no}
experimental information exists on the
$\Delta S=0$ $\frac{3}{2}^{-}   \leftrightarrow\frac{3}{2}^{+}$ or
$\frac{1}{2}^{-}\leftrightarrow\frac{1}{2}^{+}$   amplitudes.
Since we seek only to provide an estimate for $d_\Delta^\pm$
and not to perform a detailed treatment of the underlying quark
dynamics, we use the results of Ref. \cite{resonance1}
for the   $\Delta S=1$ $\frac{1}{2}^{-}\leftrightarrow\frac{1}{2}^{+}$
amplitudes   for
guidance in setting the scale of the $\Delta S=0$ weak matrix   elements,
but recognize that there can be considerable uncertainty in   these numbers

\medskip   \noindent (ii) In computing the amplitudes associated
with Figs. 2 b,c we   require the   electromagnetic (EM)
$R(\frac{1}{2}^{-})\to\Delta(1232)$   and $R(\frac{3}{2}^{-})\to N(939)$
transition amplitudes. The   EM decays of the $\frac{1}{2}^{-}$
resonances to the $\Delta(1232)$   have not been observed,
while the partial widths for $R(\frac{3}{2}^{-})\to   p\gamma$
have been seen at the expected rates. For purposes of
estimating   $d_\Delta^\pm$, then, we consider only the
contributions from Fig. 2c   involving the   $\frac{3}{2}^{-}$ resonances.

The lowest order weak and EM Lagrangians needed in computing
the amplitudes of Fig. 2c are
\begin{eqnarray}   \label{eq:resem}
{\cal L}^{RN}_\sst{EM} & = & \frac{eC_\sst{R}}{\lamchi}
{\bar   R}_\mu\gamma_\nu p   F^{\mu\nu} \   \ + {\mbox{h.c.}}\\
\label{eq:resweak}
{\cal L}^{R\Delta}_\sst{PV} & = & i {W_\sst{R}}{\bar R}^\mu\Delta_\mu \ \ +
{\mbox{h.c.}}
\end{eqnarray}
where we have omitted labels associated with charge and
isospin for   simplicity and where $R^\mu$ denotes the
spin-$3/2$ field. The constants   $C_\sst{R}$ and   $W_\sst{R}$ are
{\em a priori} unknown. Using Eqs. (\ref{eq:resem},   \ref{eq:resweak}),
we obtain
\begin{equation}
\label{eq:ddres}
d_\Delta^\pm = {C_\sst{R} W_\sst{R}\over M_R-M_\Delta} \ \ \ .
\end{equation}
The experimental EM decay widths given in \cite{pdg} imply
that   $|C_{1520}| \approx  0.98 \pm 0.05,   |C_{1700}|  \approx  0.70 \pm
0.13$
with the overall sign of each uncertain. For the weak amplitudes
$W_\sst{R}$,
we   note that the analysis of Ref. \cite{resonance1} obtained
$|W_\sst{R}(\Delta S=1)|\sim 2\times 10^{-7}$ GeV $\approx 5
g_\pi\Lambda_\chi$.
Writing our estimates for $d_\Delta^\pm$ in terms of   this
quantity we obtain
\begin{eqnarray}   \label{eq:ddres2}
d_\Delta^\pm\sim 17 g_\pi  {W_{1520}\over W_R(\Delta   S=1)} +8 g_\pi
{W_{1700}\over W_R(\Delta S=1)}
\end{eqnarray}   with an uncertainty as to the relative
and overall phase.   To the extent that
$|W_\sst{R}(\Delta S=0)|\sim |W_\sst{R}(\Delta S=1)|$,   we
would expect then $|d_\Delta^\pm|\sim (10-25) g_\pi$. By
comparison, for the   $N\to \Delta$ transition anapole moment
we obtain   $a_\Delta^{CT}(VMD)\sim   -15 g_\pi$ \cite{long}
using the \lq\lq best values" of Ref. \cite{fcdh,ddh}.

Based on NDA, one might have expected $|W_\sst{R}(\Delta S=0)|\sim
\Lambda_\chi g_\pi$
and, thus, $d_\Delta\sim g_\pi$. However, the results of   Ref.
\cite{resonance1}
give $|W_\sst{R}(\Delta S=1)|\sim 5 g_\pi\Lambda_\chi$,   while
the energy denominators in Eq. (\ref{eq:ddres}) give   additional
enhancement factors of two-to-three. Since the $\Delta S=0$   amplitudes
are generally further enhanced by $V_{ud}/V_{us}$ as well as neutral
current contributions, our estimate of $d_\Delta^\pm$   could potentially
be four to five times larger than given in Eq.   (\ref{eq:ddres2}) with
$|W_\sst{R}(\Delta S=0)|\sim |W_\sst{R}(\Delta S=1)|$. Hence,
we quote  a   \lq\lq reasonable range" $|d_\Delta^\pm|=(1-100) g_\pi$
based on this possible factor of four
enhancement. Our \lq\lq best   value" $|d_\Delta^\pm({\mbox{best}})|=25 g_\pi$
is found   by taking $|W_\sst{R}(\Delta S=0)|\sim |W_\sst{R}(\Delta   S=1)|$.
Substitution into Eq. (\ref{eq:photon1}) yields   $A_\gamma^\pm\sim
1.3\times 10^{-6}$
as a reasonable estimate for the size of the effect.

To estimate the statistical precision with which one
might measure an   asymmetry of this magnitude, we use
the standard figure of merit\cite{mus94},   along with
experimental conditions roughly appropriate for
a deuterium   target and the G0 detector at Jefferson Lab\cite{krl}:
luminosity =   0.25 Mhz/nb; $d\sigma/d\Omega_\pi dp_\pi\sim 4
$nb/(MeV$/c$-sr);
$\Delta\Omega_\pi \Delta p_\pi\sim $ 0.5 (MeV$/c$-sr); and
photon polarization   $P_\gamma\sim 0.5$. With one month of
running, one could achieve a 15\%   (statistical)
determination of $\alrd$, which would be more than
adequate to   address the physics issues considered here.

This work was supported in part under U.S. Department of
Energy contracts   \#DE-AC05-84ER40150 and \#DE-FGO2-00ER41146,
the National Science   Foundation under award PHY98-01875, and
a National   Science Foundation Young Investigator Award. CMM
acknowledges a fellowship   from FAPESP (Brazil), grant 99/00080-5.
We thank K. Gustafsson,   T. Ito, J. Martin and R. McKeown for helpful
discussions.


\begin{figure}
\vspace{-4.1cm}
\epsfxsize=9.0cm
\centerline{\epsffile{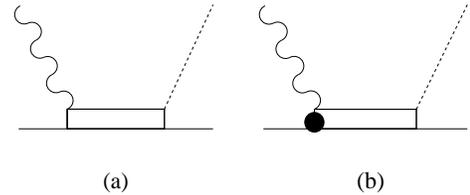}}
\vspace{-4.1cm}
\caption{Feynman diagrams for the ${\vec \gamma}N\to \Delta \to \pi   N $
process.
The dark circle indicates a parity violating coupling.}
\label{Fig.1}
\end{figure}
\begin{figure}
\vspace{-4.1cm}
\epsfxsize=9.0cm
\centerline{\epsffile{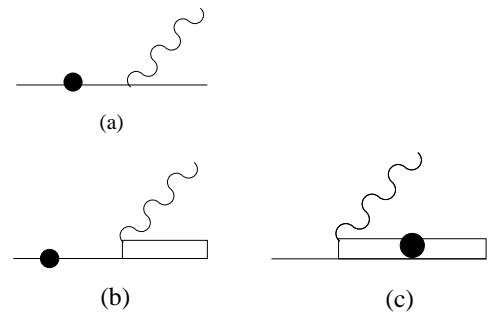}}
\vspace{-3.5cm}
\caption{Feynman diagrams for the resonance saturation model.
The dark circle indicates a parity violating coupling.
The double line denotes spin $3/2$ states.
The intermediate state $R$ has negative parity.
Diagram (a) is for hyperon weak radiative decays while (b)
and (c)   are for $d^\pm_\Delta$.}
\label{Fig.2}
\end{figure}

\end{document}